\documentclass[fleqn,usenatbib]{mnras}

\usepackage[T1]{fontenc}
\usepackage{ae,aecompl}

\usepackage{graphicx}	
\usepackage{amsmath}	
\usepackage{amssymb}	

\usepackage{hyperref}
\usepackage{cleveref}
\usepackage{booktabs}
\usepackage{multirow}
\usepackage[dvipsnames]{xcolor}
\usepackage[normalem]{ulem}

\newcommand*{\ksM}{\text{km/s Mpc$^{-1} $}}
\newcommand*{\eps}{\text{$\log_{10}(\epsilon)$}}
\newcommand*{\tauG}{\text{$\log_{10}(\tau/{\rm Gyr})$}}

\newcommand{\diff}{\mathrm{d}}

\defcitealias{Riess18_H0}{R18}
\defcitealias{Vattis19}{V19}
\defcitealias{Blackadder14}{BK14}

\title[Late-time decaying dark matter]{Late-time decaying dark matter: constraints and implications for the $H_0$-tension}

\author[B. S. Haridasu et al.]{
Balakrishna S. Haridasu,$^{1,2}$\thanks{haridasu@roma2.infn.it}
Matteo Viel,$^{3,4,5,6}$\thanks{viel@sissa.it}\\
$^{1}$Dipartimento di Fisica, Universit\`a di Roma "Tor Vergata", Via della Ricerca Scientifica 1, I-00133, Roma, Italy\\
$^{2}$Sezione INFN, Universit\`a di Roma "Tor Vergata", Via della Ricerca Scientifica 1, I-00133, Roma, Italy\\
$^{3}$SISSA-International School for Advanced Studies, Via Bonomea 265, 34136 Trieste, Italy\\
$^{4}$INFN, Sezione di Trieste, Via Valerio 2, I-34127 Trieste, Italy\\
$^{5}$INAF - Osservatorio Astronomico di Trieste, Via G. B. Tiepolo 11, I-34143 Trieste, Italy\\
$^{6}$IFPU, Institute for Fundamental Physics of the Universe, via Beirut 2, 34151 Trieste, Italy
}

\date{Accepted XXX. Received YYY; in original form ZZZ}

\pubyear{2020}

\begin{document}
\label{firstpage}
\pagerange{\pageref{firstpage}--\pageref{lastpage}}
\maketitle

\begin{abstract}
We constrain and update the bounds on the life-time of a decaying dark matter model with a warm massive daughter particle using the most recent low-redshift probes. We use Supernovae Type-Ia, Baryon Acoustic Oscillations and the time delay measurements of gravitationally lensed quasars. These data sets are complemented by the early universe priors taken from the Cosmic Microwave background. For the maximum allowed fraction of the relativistic daughter particle, the updated bounds on the life-time are found to be $\tau > 9\, \rm{Gyr}$ and $\tau >11\,\rm{Gyr}$ at $95\%$ C.L., for the two-body and many-body decay scenarios, respectively. We also comment on the recent proposal that the current two-body decaying dark matter model can provide resolution for the $H_0$-tension, by contrasting against the standard $\Lambda$CDM model. We infer that the current dark matter decaying scenario is unlikely to alleviate the $H_0$-tension. We find that the decaying dark matter is able to reduce the trend of the decreasing $H_0$ values with increasing lens redshifts observed in the strong lensing dataset.
\end{abstract}

\begin{keywords}
(\textit{cosmology}:) dark matter --  cosmological parameters 
\end{keywords}



\section{Introduction}

A decaying dark matter particle provides interesting phenomenological aspects predicting a variation in the late-time evolution of the universe, compared to a standard cold dark matter scenario. Several works \citep{Audren14a, Blackadder14, Poulin16}, have used cosmological data to constrain the decay characteristics of such a dark matter candidate, putting limits on the life-times of the parent particle. In \citet{Blackadder14} (hereafter \citetalias{Blackadder14}) and in \citet{Blackadder15} a dark matter decay scenario has been developed where the massive daughter particle is not necessarily cold at the epoch  of decay and hence provides a dynamical equation of state for the collective dark matter behaviour. Here, we implement this model to constrain the decay characteristics with the most recent low-redshift cosmological data. This model has been earlier constrained against the Supernovae type-Ia datasets (Union2.1 compilation taken from \citet{Suzuki12}) and using the high-redshift Cosmic Microwave Background (CMB) priors from \textit{Planck} 2013 \citet{Ade13} release, in \citetalias{Blackadder14}.

Several implementations of decaying dark matter scenarios are interesting and  possibly complementary to current scenario: i) decaying dark matter resulting in effective neutrino density \citep{Hasenkamp12}; ii) a fraction of initial dark matter decaying into radiation \citep{Aubourg14, Audren14a}, yielding a limit of $\tau > 150 \, \rm{Gyr}$ using CMB data in \citet{Poulin16}; iii) dark matter decay injecting energy to the bayronic gas component \citep{Zhang07}; iv) dark matter decaying to neutrinos scenario  with $95\%$ C.L. of $\tau> 700 \, \rm{Gyr}$ and $\tau> 100 \, \rm{Gyr}$, respectively reported in \citet{Gong08} and \citet{DeLopeAmigo09}. The effects of decaying dark matter on the structure formation were studied in several works like \citet{Wang12}, assessing sensitivities of the life-times to the kick velocities, reporting that Euclid \citep{Amendola16}, LSST \citep{Mandelbaum18} surveys can be sensitive to $\tau \lesssim 5 \, \rm{Gyr}$ for kick velocities $<90 \, \rm{km/s}$. In \citet{Wang13} an upper limit of $\tau \lesssim 10 \, \rm{Gyr}$ was placed for kick velocities of $30-70 \, \rm{km/s}$ using Ly$-\alpha$ forest data \citep{Kim04, McDonald06}. Some approaches also addressed dynamical dark matter scenarios with time varying equation of state, due to interacting ensemble of unstable dark matter particles decaying into ordinary matter \citep{Dienes11a, Dienes11b}. Recently, one such implementation with number of variable degrees of freedom corresponding to unstable decay particles \citep{Desai19} has utilised SN data to constrain the decay characteristics, in effect considering decay ensembles to radiation alone. 

On the other hand, owing to the well established $H_0$-tension \citep{Riess19_Nat, Bernal16, Feeney17, Addison17}, now reaching $\sim 5\sigma$ level as reported in \citet{Wong19}, several propositions have been put forward to potentially address the growing crisis. Several of these proposals focus on the early-universe modifications such as early dark energy \citep{Poulin18, Ye20}  and vacuum phase transitions \citep{DiValentino:2017rcr}, interacting dark energy \citet{Pan19,DiValentino:2017iww} and other scenarios \citet{Banihashemi:2018has, Valentino18, Raveri17}. Alternatively, some approaches focus on the modification of the local estimate \citep{Hoscheit17, Shanks19, Schoneberg19} (see also \citet{Kenworthy19, Lukovic19}). In this context, \citet{Vattis19} (hereafter \citetalias{Vattis19}) have recently proposed that the current decaying dark matter model with a warm massive decay particle can possibly alleviate the $H_0$-tension, by performing a simple analysis on the expansion rate data.
Here, we exploit the opportunity to also revisit the claim with more new data: the $\sim 1050$ Supernovae Type-Ia (SN) compilation in \citet{Scolnic17} and the gravitationally lensed quasar time delay (SL) measurements \citet{Wong19} and a compilation of up-to-date Baryon Acoustic Oscillations (BAO) datasets. We complement these low-redshift probes with the CMB priors as suggested in \citet{Verde17}, which indeed is a more apt way of imposing priors at recombination epoch for the late-time decaying dark matter model. Several other works \citep{Enqvist15, Bringmann18, Pandey19, Xiao19} have also considered decaying dark matter as a means to alleviate the $H_0$-tension. Alongside the decaying scenarios, various other modifications to the dark matter sector in general have also been explored in the context of $H_0$-tension \citep{Ko17, Raveri17a, daSilva18, DEramo18, BuenAbad18, Choi20, Blinov20}, most of which are shown to reduce the significance of the tension. 

The authors of \citetalias{Vattis19} clearly state that the usual tendency of the decaying dark matter models would be to reduce the expansion rate at late-times in comparison to the early time expectation. Thus, they suggest that a specific combination of the decay characteristics can bring the expansion rate at $z = 0$ in agreement with the evolution of $H(z)$ at higher redshifts as measured at recombination epoch. This in-turn is one of the motivations, as we intend to assess the decay scenario with additional data to validate the claim, as they have only used the expansion rate information from a BAO compilation. Note that using only the expansion rate data would be loosely constraining and can be elusive to the well-constrained angular scales at the recombination epoch. 

The organisation of the paper is as follows: In \Cref{sec:Theory} we describe the theoretical model and data analysis implemented, followed by the results and discussion in \Cref{sec:Results}, and concluding remarks in \Cref{sec:Conclusions}. 

\section{Modelling and Analysis}
\label{sec:Theory}
We implement the decaying dark matter (here after $\Lambda$DDM) formalism with a possibly warm/relativistic daughter particle, essentially following the formalism developed in \cite{Blackadder14, Blackadder15}, where the two-body decay of the parent dark matter particle produces a heavy daughter particle which at creation could be relativistic (warm), eventually becoming non-relativistic and a second massless relativistic particle. This scenario produces a range of possibilities with the decay rate and fraction of parent energy density split amongst the two daughter particles. Due to this diversity in the time of decay and the fractional energy densities transferred to the daughter particles the system of equations depending on the expansion rate needs to be solved in an iterative fashion, to provide the the final expansion history. In this section, we describe the model where the expansion rate has to be inferred simultaneously assessing the respective energy densities of the parent, and the two daughter particles. We keep the description of the model brief and refer to the original work in \citet{Blackadder14, Blackadder15}, for further details.

Within the two-body decay system, the evolution of the parent and the massless daughter particle in terms of the scale factor ($a$) can be written as,
  \begin{equation}
 \label{eqn:decay_pandmld}
 \begin{aligned}
 \frac{d \rho_0}{d t} + 3 \frac{\dot{a}}{a} \rho_0 &= -\Gamma \rho_0 \\
 \frac{d \rho_1}{d t} + 4 \frac{\dot{a}}{a} \rho_1 &= - \epsilon \Gamma \rho_0,
 \end{aligned}
 \end{equation} 
respectively. Here the decay rate of the parent particle, $\Gamma \equiv 1/\tau$ ($\tau$ being the lifetime) and the fraction ($\epsilon$) of the rest mass energy acquired by the massless relativistic particle through the decay are the two decay parameters. As for the massive daughter particle, the energy density at a particular instance in evolution has to be averaged over all the decays that have taken place thus far, also accounting for their dynamic equation of state (EoS). The massive daughter particles might (are allowed to) be relativistic at the time of decay ($a_{\rm D}$), and can indeed exhibit varied behaviour depending on weather $a_{\rm D} \ll 1$, early decay that gets redshifted or $a_{\rm D} \sim 1$, a late-time decay. As elaborated in \citetalias{Blackadder14}, taking into account all the aforementioned effects the energy density of the massive particle can be written as, 

\begin{equation}
\label{eqn:decay_md}
    \rho_2(a) = \frac{\mathcal{A}}{a^{3}}\int_{a_{*}}^{a}\frac{e^{-\Gamma t(a_{\rm D})}}{a_{\rm D}H(a_{\rm D})} \left[\frac{\epsilon^2}{1-2\epsilon}\left(\frac{a_{\rm D}}{a}\right)^2 + 1 \right]^{1/2}\diff a_{\rm D},
\end{equation}
where $a_{*}$ is scale factor corresponding to the recombination and the normalisation factor $\mathcal{A}$. The expansion history finally is given as the summation of the energy densities of all the contributing components, 

\begin{equation}
    \label{eqn:Hofz}
    H^2(a) = \frac{8\pi G}{3}\left[\rho_0(a)+\rho_1(a)+\rho_2(a)+\rho_{\rm b}(a) + \rho_{\rm r}(a)\right] + \frac{1}{3}\Lambda c^2
\end{equation}
where, $\sum_{i=0}^{2} \rho_{\rm{i}}(a)$ corresponds to the total contribution of decaying dark matter components, $\rho_{\rm b}\,, \rho_{\rm r}$ are the contributions of baryons and radiation\footnote{Here radiation includes both the contributions of photons and neutrinos, which we implement as in \citetalias{Blackadder14, Vattis19} following \citet{Komatsu11}, however they affect the late-time dynamics minimally.}, respectively. The initial conditions for the decay particles are set such that at $ \rho_1(a_{*}) = \rho_2(a_{*}) = 0$ and $\rho_0(a_{*})$ comprises the entire dark matter contribution at $a_{*}$. 

In conjunction to the two-body decay, in \citetalias{Blackadder14} also a many-body decay scenario is developed, where the  massive daughter particle is set to be cold. This formalism is equivalent and draws parallels to the one implemented in \citet{Aubourg_2015, Audren14a, Poulin16}, where the whole parent particle energy density is allowed to decay. In the many-body  scenario the energy density of the massive daughter particle is much simpler to estimate in a similar way as the parent and massless daughter particle \Cref{eqn:decay_pandmld} and the \Cref{eqn:decay_md} are replaced by,

\begin{equation}
 \label{eqn:decay_md_MB}
 \begin{aligned}
 \frac{d \rho_2}{d t} &= (1 - \epsilon) \Gamma \rho_0 -  3 \frac{\dot{a}}{a} \rho_2\,\\
 \rho_2(a) &= \frac{\mathcal{A} (1-\epsilon)}{a^{3}} \left[e^{-\Gamma t(a_{*})} - e^{-\Gamma t(a)}\right].
 \end{aligned}
 \end{equation}

The corresponding distances are estimated as $D_{\rm L}(z) = c(1+z)\int_{0}^{z} \diff\xi/H(\xi)$, once the $H(\xi)$ is obtained iteratively. 
As for the analysis, we follow the same procedure as described in \cite{Vattis19}, however using more low-redshift datasets: the most recent Supernovae Type-Ia (SN) compilation in \citet{Scolnic17}, an up-to-date compilation of BAOs \footnote{We use the estimates of the comoving angular diameter distance $D_{\rm A}(z)/r_d$ and the Hubble rate $H(z)\times r_d$ provided at $z = \{0.38, 0.51, 0.61\}$ by \cite{Alam16}, which combines the analysis of different companion works on SDSS DR-12, in a consensus result. At intermediate redshifts we utilised the more recent measurements provided by SDSS-IV eBOSS data release \cite{Zhao18_dr14}, at redshifts $z=\{0.98,1.230,1.526, 1.944\}$. Finally the farthest measurements in redshift are provided by the auto-correlation of the Ly-$\alpha$ forest and the cross-correlation of Ly-$\alpha$ and quasars at $z\sim 2.4$ in \cite{Blomqvist19} and \cite{SainteAgathe19}.} observables and the 6 gravitationally lensed quasar time delay (SL) dataset presented in \citet{Wong19}, which provides the $H_0$ measurements in our analysis. To set the initial conditions for the dark matter and baryon densities we use early-time priors suggested in \citet{Verde17}\footnote{We acknowledge the authors of \citet{Verde17}, for providing us with the covariance matrix of the observables. Please refer to Table 2 presented therein. }, where the corresponding energy densities (physical) and expansion rate at recombination ($a_{*} = 1089$) are constrained, disentangling the late-time physics. These priors indeed complement very-well the low-redshift probes to test the late-time effects of decaying dark matter, while having early universe physics unchanged from $\Lambda$CDM. Setting the initial conditions enforces that no dark matter decays have taken place before the recombination epoch.

The parameters of the model sampled upon in the MCMC analysis are the matter densities for initial dark matter, baryons $\Omega_{\rm DM}^{*}$, $\Omega_{\rm b}$, and the $H_0^{*}$ corresponding to the early-time $\Lambda$CDM model `fixed' at recombination epoch ($a_{*}$), accompanied by the two decay parameters: decay rate of the parent particle, $\Gamma \equiv 1/\tau$ ($\tau$ being the lifetime) and the fraction ($\epsilon$) of the rest mass energy acquired by the massless relativistic particle through the decay. These early-time parameters are utilised to compute the sound horizon at drag epoch ($r_d$), through the fitting formula provided in \citet{Aubourg14}. This ensures that the early-time scale of the sound horizon at drag epoch is not affected by the late-time decaying of the dark matter density. The actual dark matter density (along with massless daughter) today and present expansion rate would be given by the iteratively computed $\sum_{i=0}^{2} \rho_{\rm{i}}(a=1)$ and the $H(a=1)$ in \Cref{eqn:Hofz}, respectively. While for the $\Lambda$CDM model $\Omega_{\rm DM}^{*}, H_0^{*}$ would retain the standard definition. To sample larger ranges of the parameter space, the decay parameters are sampled in logarithmic scales in the ranges of $-4 \leq \log_{10}(\Gamma)\leq 3 $ and $-4 \leq \eps < \log_{10}(0.5)$. Note that these logarithmic flat priors are the same as in \citetalias{Vattis19}.

We implement a simple Bayesian analysis using the \texttt{emcee}\footnote{\href{http://dfm.io/emcee/current/}{http://dfm.io/emcee/current/}} \citep{Foreman-Mackey13} package to perform the analysis and the \texttt{getdist}\footnote{\href{https://getdist.readthedocs.io/}{https://getdist.readthedocs.io/}} package \cite{Lewis19} to analyse the posteriors.  We also use the \texttt{ChainConsumer} package \citep{Hinton16}, publicly available\footnote{\href{https://github.com/Samreay/ChainConsumer/tree/Final-Paper}{https://github.com/Samreay/ChainConsumer/tree/Final-Paper}.}.

\section{Results}
\label{sec:Results}

{\renewcommand{\arraystretch}{1.5}%
    \setlength{\tabcolsep}{7pt}%
    \begin{table*}
        \centering
        \caption{Constraints in the $\Lambda$DDM model at 68\% confidence level obtained with and without the inclusion of BAO datasets. We quote the maximum posterior and the $16^{\rm th} , 84^{\rm th}$ percentiles as the uncertainty. We also report the best-fit (b.f) for the $\Lambda$DDM model which differs from the max-posterior, when the BAO data is included. Here $H_0^*$ is reported in the units $\ksM$.}
        \label{tab:LDDM_params}
        \begin{tabular}{c|ccc|ccc}
        \toprule
        Data &\multicolumn{3}{c|}{SN+SL}&\multicolumn{3}{c}{SN+SL+BAO}\\
        \hline
        Model & \multicolumn{2}{c}{$\Lambda$DDM}& \multicolumn{1}{c|}{$\Lambda$CDM}& \multicolumn{2}{c}{$\Lambda$DDM}& \multicolumn{1}{c}{$\Lambda$CDM}\\
              & b.f & $1\sigma$ & $1\sigma$ & b.f & $1\sigma$ & $1\sigma$ \\
		\hline
		$\Omega_{\rm DM}^{*}$ & $0.229$ & $ 0.228^{+0.011}_{-0.010}$ & $ 0.227^{+0.012}_{-0.010}$ & $0.245$ & $ 0.249^{+0.006}_{-0.006}$ & $ 0.251^{+0.005}_{-0.006}$ \\ 
		$\Omega_{\rm b}\times 10^{2}$ & $4.3$ & $ 4.29^{+0.20}_{-0.19} $ & $ 4.24^{+0.23}_{-0.16}$ & $4.6$ &$ 4.71.0^{+0.09}_{-0.11}$ & $ 4.70^{+0.10}_{-0.09}$ \\ 
		$H_0^{*}$ & $72.01$ &$72.1^{+1.6}_{-1.7}$ & $72.2^{+1.6}_{-1.7}$ & $69.64$ & $68.90^{+0.78}_{-0.75}$ & $68.98^{+0.57}_{-0.81}$ \\  
		\bottomrule
    \end{tabular}
    \end{table*}
}

We firstly discuss the constraints on energy densities and limits on the decay parameters and finally comment on inferences for $H_0$-tension, within the current decaying dark matter scenario. In \Cref{tab:LDDM_params} we report the $68\%$ C.L. limits on the $\Omega_{\rm DM}^{*}$ and $H_0^{*}$. The dataset combination of SN+SL as expected provides larger values of $H_0$, allowing for lower values of both the dark matter and baryon energy densities, with no distinguishable difference between the two decaying scenarios and $\Lambda$CDM. With the inclusion of the BAO dataset the energy densities are pushed towards higher values, yielding a low value of $H_0^*$. Marginalising on the decay parameters, the many-body and two-body decay scenarios do not show any discernible difference for the marginalised constraints on dark matter energy density and $H_0$. The best-fit value of $H_0^* = 69.92 \,\ksM$ using the SN+SL+BAO data in the many-body scenario is also comparable to the  $H_0^* = 69.64\, \ksM $ for two-body case presented in \cref{tab:LDDM_params}.

\begin{figure}
    \centering
    \includegraphics[scale=0.35]{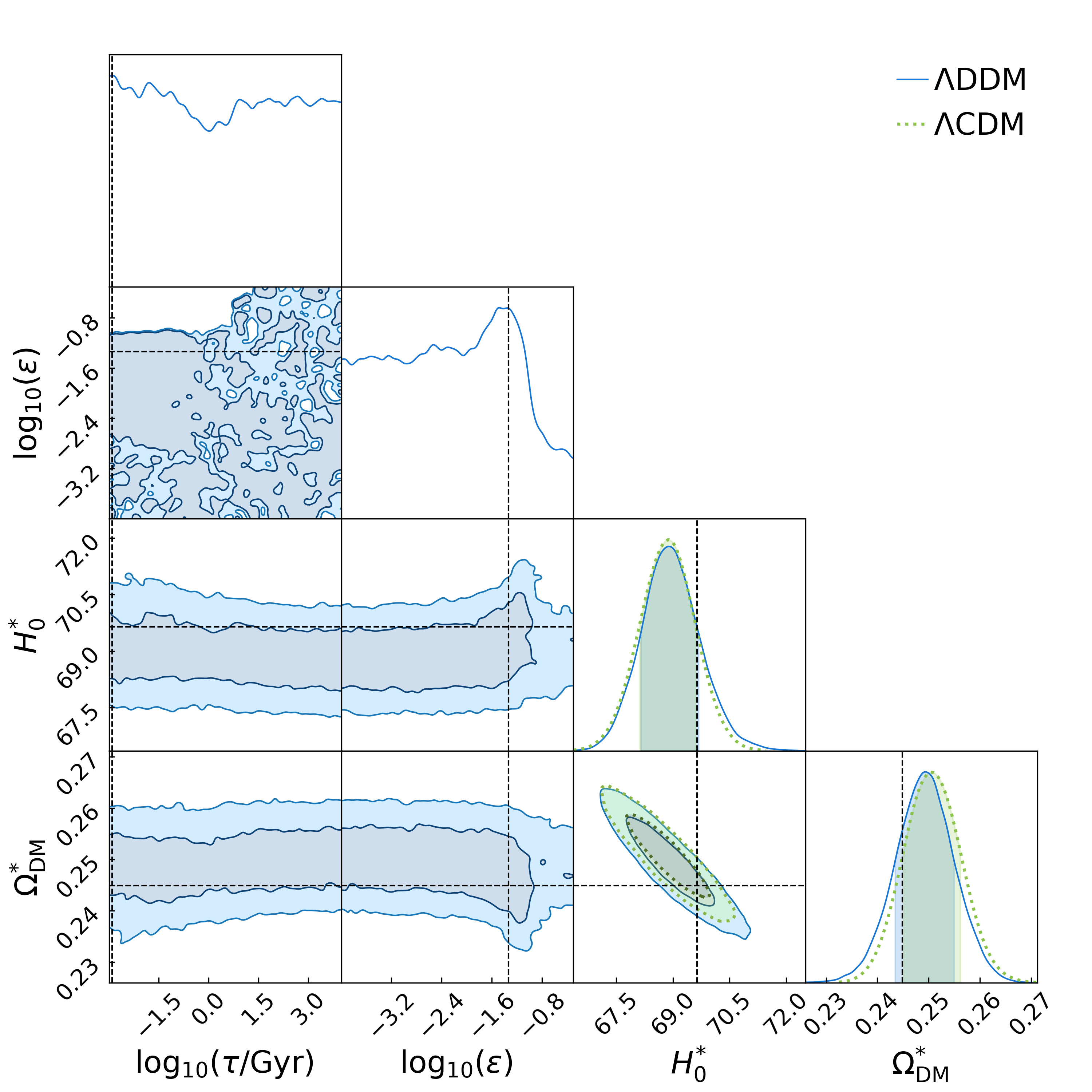}
    \caption{Here we compare the constraints in the $\Lambda$DDM model with two-body decay to the $\Lambda$CDM model, using the SN+SL+BAO dataset. The dashed line in each of the panel corresponds to the best-fit parameters reported in \Cref{tab:LDDM_params}. The inner and outer contours correspond to $68\%\, 95\%$, confidence levels, respectively. $H_0^*$ is in the units of $\ksM$.}
    \label{fig:LDDM_CON}
\end{figure}

When including the BAO dataset, both for the $\Lambda$DDM and $\Lambda$CDM models, the constraints on $\Omega_{\rm DM}^{*}$ and $H_0^{*}$, shift to higher and lower values, respectively. For the $\Lambda$CDM model, we find a value of $H_0 = 68.98^{+0.57}_{-0.81} \ksM$, which is in very good agreement with the earlier reported $H_0$ values inverse distance ladder analysis \citep{Lemos:2018smw, Camarena19}, especially with \citet{Lyu20} where the SL dataset is taken into account. While the mean values of $H_0^*$ are in good agreement between $\Lambda$CDM and the two-body $\Lambda$DDM models, we find that the best-fit of the latter model has an higher value of $H_0^* = 69.64 \, \ksM$, which is at $1\sigma$ of the posterior distribution. As can be seen in \Cref{fig:LDDM_CON}, the distortion of the contours in the $\eps\, \rm{vs.}\, \Omega_{\rm DM}^{*}$ and $\eps\, \rm{vs.}\, H_{0}^{*}$ parameter space accommodates the best-fit value and is similar to the constraints presented in \citetalias{Vattis19}. We discuss the implications for $H_0$-tension later in \Cref{sec:Results_H0}. The life-time of the parent particle $\tauG$ remains unconstrained and the fraction of relativistic daughter particle, $\eps$ shows a mild peak in the 1D marginalised distribution around the best-fit value. This feature is however not noticed when BAO data is not included and both the decay parameters remain unconstrained. 

\begin{figure*}
    \centering
    \includegraphics[scale=0.5]{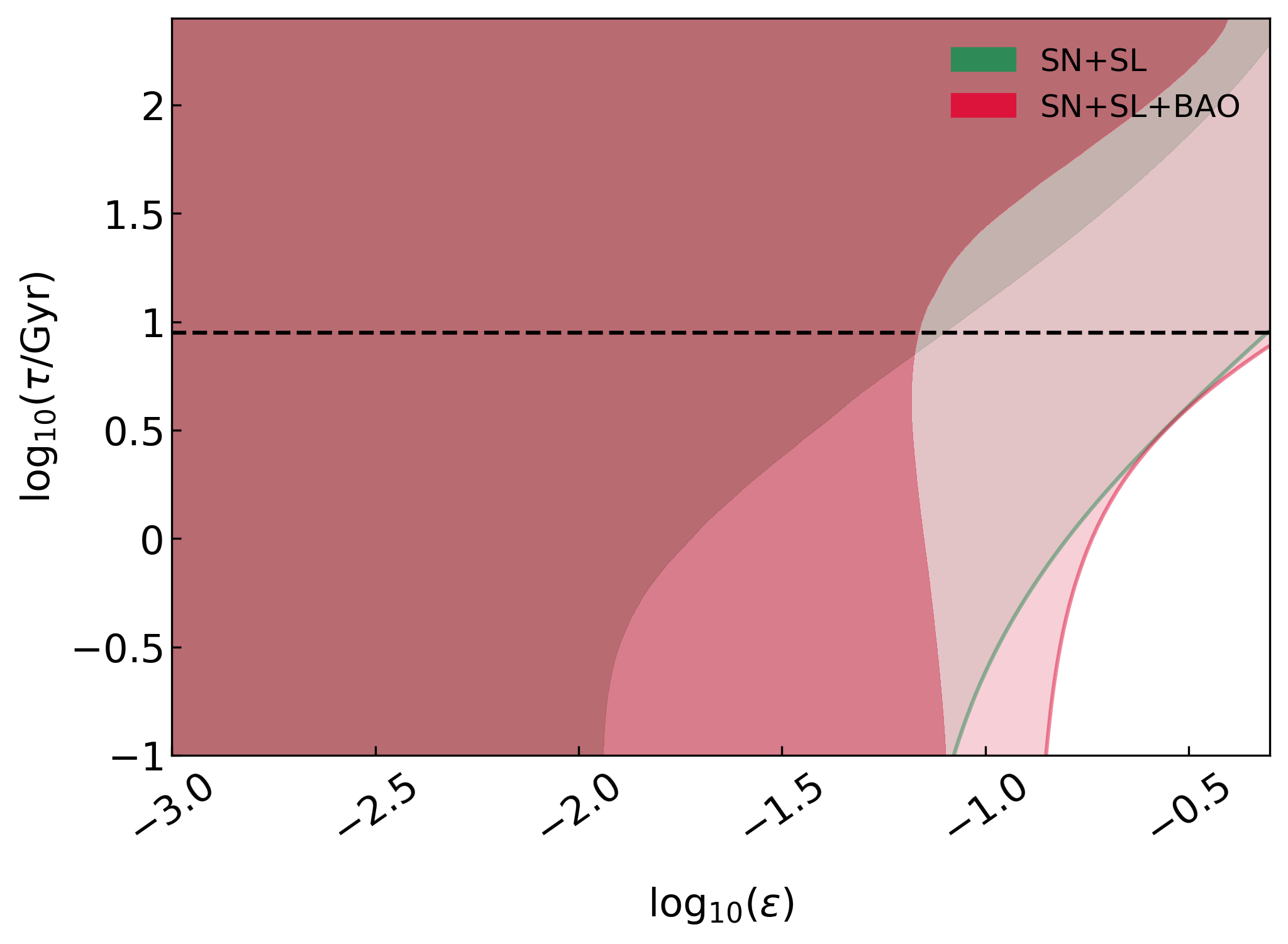}
    \includegraphics[scale=0.5]{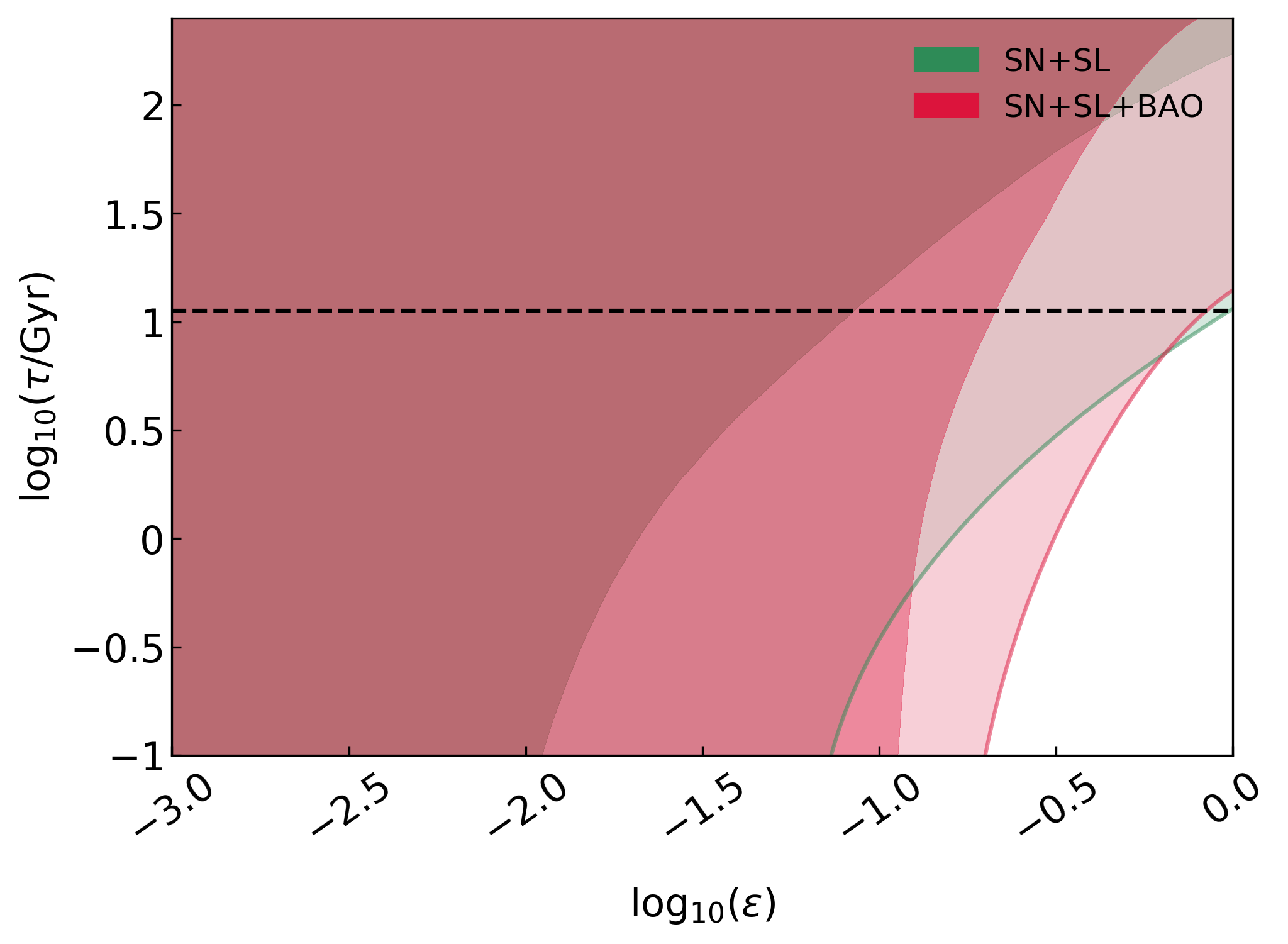}
    \caption{\textit{Left}: Constraints for the $\eps \,, \tauG $ parameter space for the two-body $\Lambda$DDM model reported for SN+SL (green) and SN+SL+BAO (red). The contours depict $68\%$ and $95\%$ C.L. limits, respectively. \textit{Right}: Same as left panel, but for the many-body decay scenario. }
    \label{fig:LDDM_decay}
\end{figure*}

In \Cref{fig:LDDM_decay}, we show the constraints on the decay parameters of the $\Lambda$DDM model. We recover the general features of the constraints presented in \citetalias{Blackadder14} and for ease of comparison, present the constraints in terms of $\tauG$, instead of the sampled $\log_{10}(\Gamma)$ parameter. Notice that our constraints from the MCMC analysis performed here, are much less stringent\footnote{We speculate that a major reason for this could be due to the Frequentist approach in \citetalias{Blackadder14}, where the rest of the parameters are fixed to there best-fit values and confidence levels are placed through $\chi^2$ cuts of the likelihood corresponding to a Gaussian-like distribution. The definition of their `goodness of fit confidence' is also  based on the reduced $\chi^2$ values.} in comparison to the ones presented in \citetalias{Blackadder14}, even with the improved SN dataset and other low-redshift probes and updated priors from the high-redshift CMB. While the inclusion of BAO data mildly strengthens the $68\%$ C.L. limits on life-time for $\eps\gtrsim -0.1$, the $95\%$ C.L. limits mostly coincide with the those obtained from the SN+SL data alone. On the contrary, for $\tauG \lesssim 0.5$, the  limits  on the allowed fraction $\eps$ are much broader than those obtained with the SN+SL data. This enlarged parameter space corresponds to the mild peak that can be noticed in the $\eps\, \rm{vs.}\, \Omega_{\rm DM}^{*}$ parameter space in \Cref{fig:LDDM_CON}, and is the case for both the two-body and many-body decay scenarios. {This clearly indicates that it is not obvious to expect that the bounds on the life-time will become stringent with the inclusion of more data. For a given $\epsilon$, the allowed range of decay life-time depends on the initial dark matter density ($\Omega_{\rm DM}^{*}$) and the late-time total energy density of all the dark matter components constrained by the low redshift data.} This highlights the importance of the BAO data in constraining the decay characteristics of the dark matter model in consideration.


\begin{figure*}
    \centering
    \includegraphics[scale=0.49]{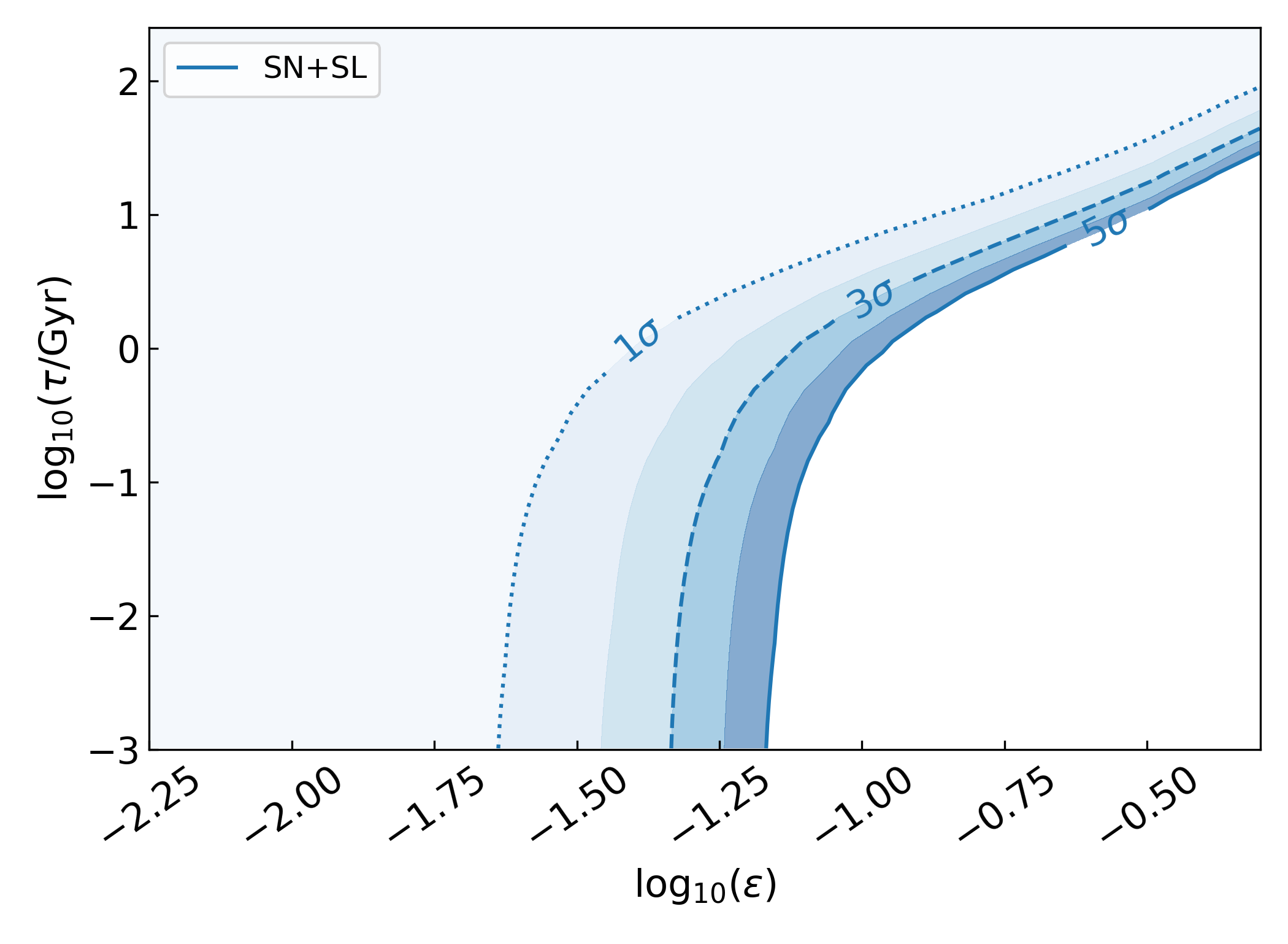}
    \includegraphics[scale=0.49]{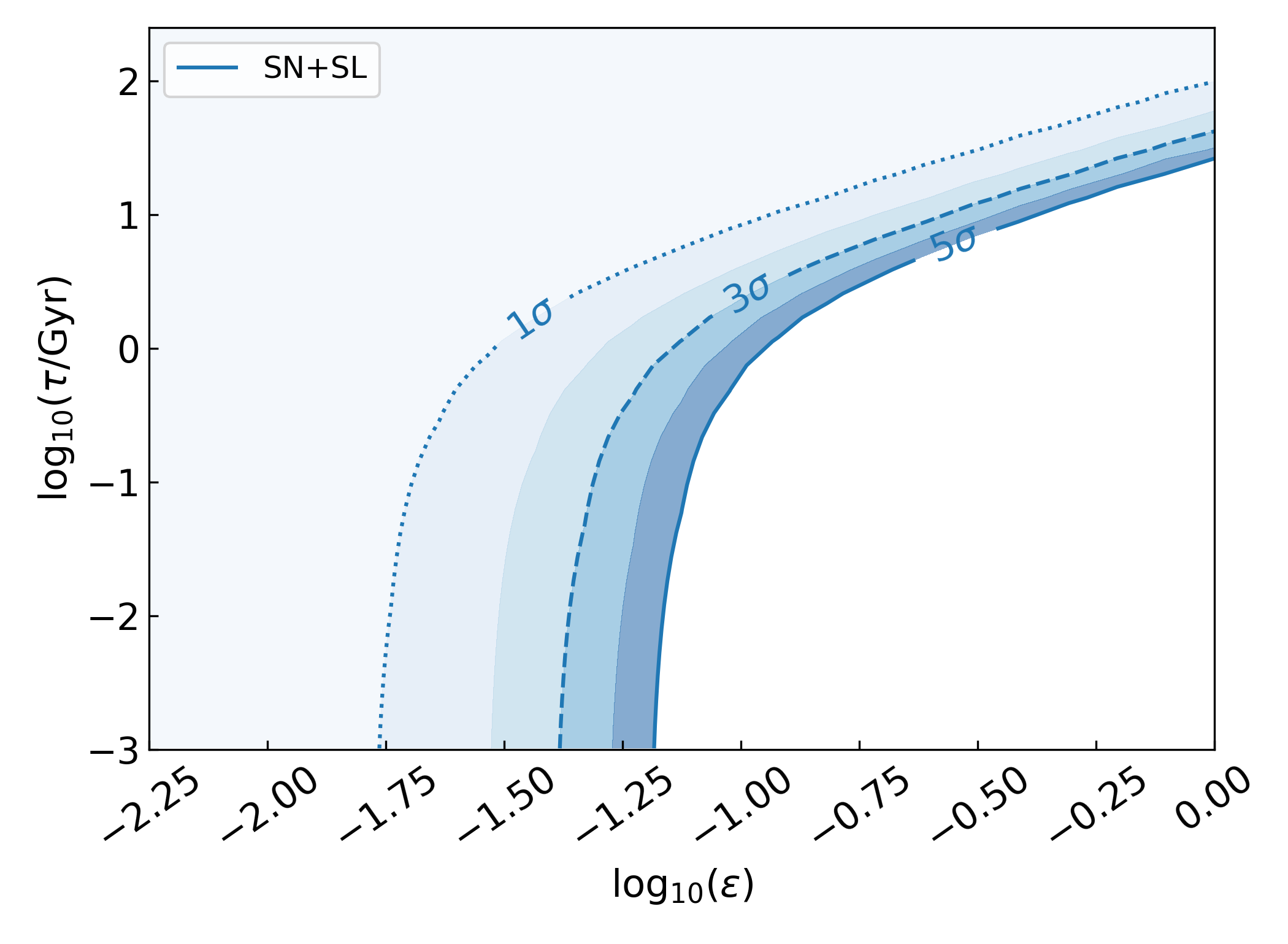}
    \hfill
    \includegraphics[scale=0.49]{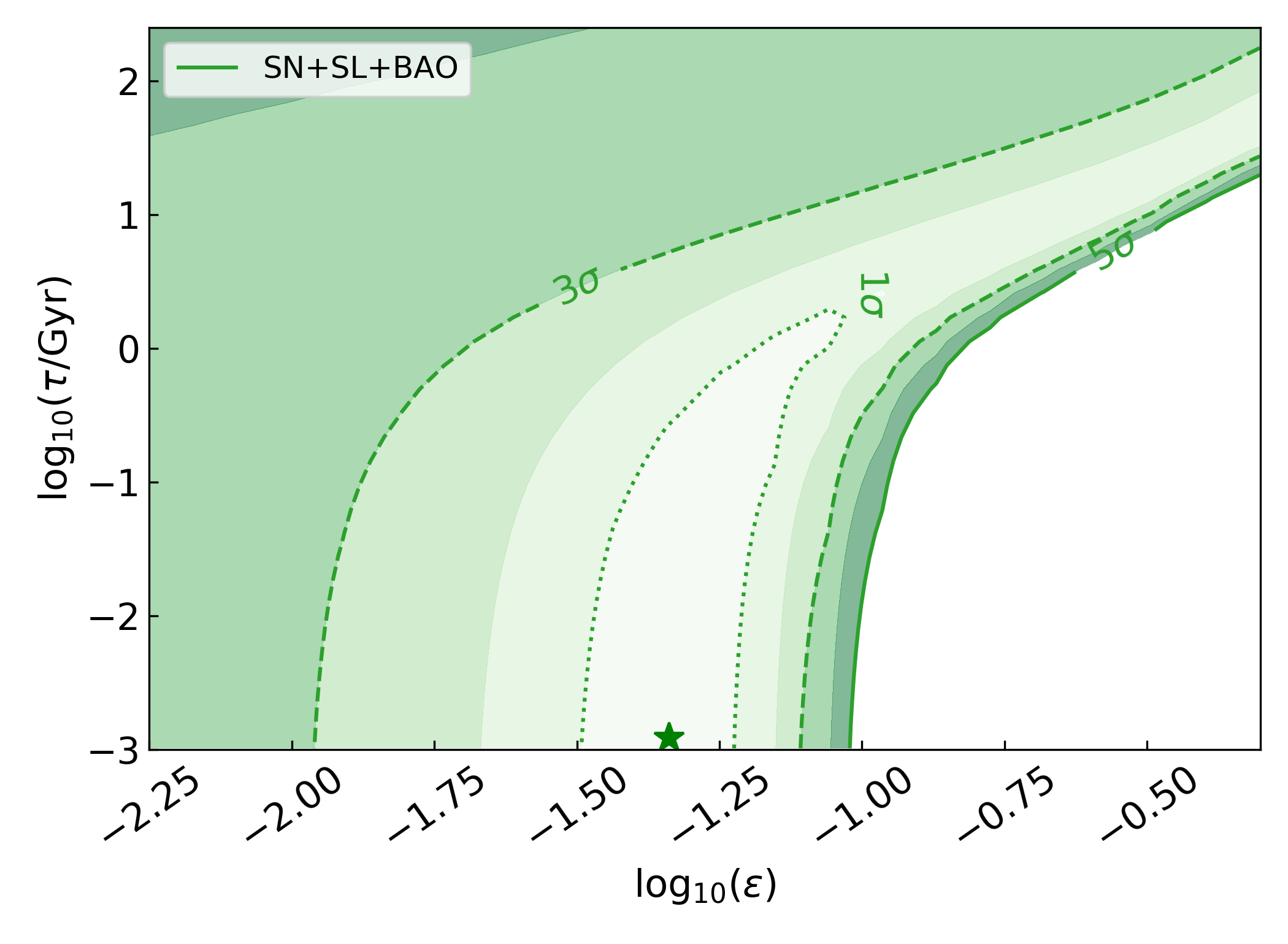}
    \includegraphics[scale=0.49]{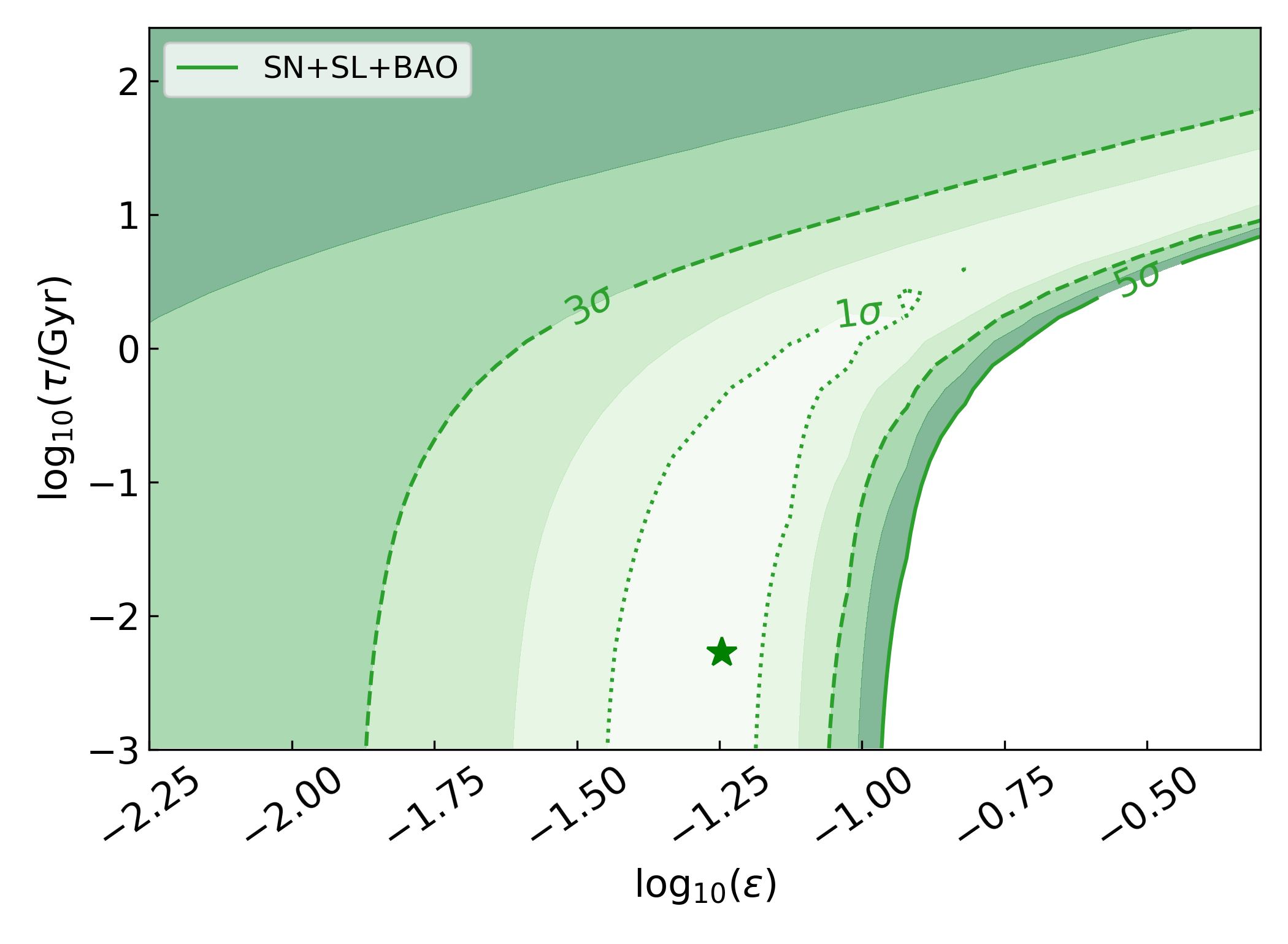}
    \centering
    \caption{\textit{Left}: Profile likelihoods for the $\eps \,, \tauG $ parameter space for the two-body $\Lambda$DDM model using the SN+SL dataset (\textit{Top}) and SN+SL+BAO (\textit{Bottom}). We show contours $1\sigma $ through $5\sigma$. \textit{Right}: Same as left panel, but for the many-body decay scenario. In the \textit{Bottom} panels star marks the best-fit values. Please note the difference in the range of the axes when comparing with \Cref{fig:LDDM_decay}.}
    \label{fig:LDDM_decay_F}
\end{figure*}

In the case of two-body decay, we find the lower limit on the life-time of particle for $\epsilon = 0.5$, to be $\tau \gtrsim 9.0\, \rm{Gyr}$ at $95\%$ C.L., which is equivalent both with and without the inclusion of BAO dataset to the SN+SL data. Similarly, for the many-body decay we find $\tau \gtrsim 11.2\, \rm{Gyr}$ at $95\%$ C.L., which is comparable to the limit of $\tau > 28\, \rm{Gyr}$ reported in \citet{Aubourg14}. Our limits are indeed less stringent in comparison to the $\tau \gtrsim 150 \times f_{\rm dcdm} \, \rm{Gyr}$, set in \citet{Poulin16}\footnote{We also verify that this limit changes only mildly with updated \textit{Planck} 2018 \citep{Planck18_parameters} dataset, to $\tau \gtrsim 154 \, \rm{Gyr}$ for $f_{\rm dcdm} = 1$.}, where $f_{\rm dcdm}$ is the fraction of initial cold dark matter that is allowed to decay. The range of lifetimes explored in our analysis span the three different ranges classified in \citet{Poulin16} as short, intermediate and long. For short lifetimes of $\tauG \lesssim -1$, we find at $95\%$ C.L., that no more than $\sim 8\%$ of the parent particle can decay to massless relativistic daughter particle and confidence regions show no preference for the life-times, similar to the inference made in \citetalias{Blackadder14}.

Noticing that the posterior in the $\eps$ {vs.} $ \tauG$ parameter space exhibits a steep cut-off, we also compute the profile likelihoods\footnote{We compute the profile likelihoods \citep{Trotta17} presented in \Cref{fig:LDDM_decay_F} by estimating the $1\sigma$ through $5\sigma$ confidence regions defined for 2-dimensional  $\Delta\chi^2 = \{2.30,\, 6.18,\, 11.83,\, 19.33,\, 28.74\}$ cuts of the likelihood, w.r.t the corresponding $\chi^2_{\rm b.f}$.} for both the dataset combinations, while fixing the rest of the parameter to their respective best-fit values (reported in \Cref{tab:LDDM_params}). As can be seen in \Cref{fig:LDDM_decay_F}, even the $5\sigma$ limits are much tighter in comparison to the $95 \%$ marginalised confidence regions in \Cref{fig:LDDM_decay}, while the $1\sigma$ limits are more relaxed demonstrating the steep behaviour of the likelihood. This essentially excludes the region of the parameter space where large fractions of parent particle quickly decaying to the relativistic massless particle, and is effectively constrained only by the low-redshift data. At the same time, we find the $5\sigma$ limits to be equivalent to those set in \citetalias{Blackadder14} and our inferences overall agree. We also recover their inference, that for very small fractions of the relativistic massless particle the contours are essentially vertical, unable to distinguish between shorter life-times. Owing to their tighter $3\sigma$ limits they were able to provide $95\%$ C.L. constraints on lifetime $\tau > 10 \, \rm{Gyr}$ for $1\%$ relativistic daughter fraction, which we are unable to place in the current analysis and this might also be due to a difference in defining the $3\sigma$ C.L . Note that this limit in \citetalias{Blackadder14}, is placed with a fixed $H_0 = 67.15\, \ksM$ and as shown in this work by comparing the marginalised confidence regions and profile likelihoods, the final inference can be affected by fixing the background parameters. Tentatively their $95\%$ C.L. limits for maximum allowed fraction of relativistic daughter particle ($\epsilon$) would be more stringent than $\tau \gtrsim 10^3 \,\rm{Gyr}$ (see Fig. 4-5 therein). Following similar procedure, we obtain $95\%$ C.L. limits of $\tau \gtrsim 59 \, \rm{Gyr} $, from SN+SL+BAO data combination for maximum allowed $\epsilon$, for a fixed best-fit value of $H_0 = 69.64\, \ksM$. This estimate is equivalent for both two-body and many-body decay scenarios and is also in good agreement with the earlier mentioned limit of $\tau > 28 \, \rm{Gyr}$ in \citet{Aubourg14}, also where $H_0 = 68\, \ksM$ is fixed. While our estimate here is clearly an improvement, being more stringent, we however choose to report our final inference from the marginalised confidence regions discussed earlier, which is seemingly less stringent, but more accurate.

It is very well expected that in the current scenario, marginalised confidence regions can be less stringent than the profile likelihoods with fixed parameters. And interestingly, we find that the larger values of $H_0^*$ (lower values of $\Omega_{\rm DM}^*$) in the MCMC sampling are mostly aligned along the bounds (contours) in the $\tauG\, \rm{vs.} \eps$ parameter space, which allows for the extended confidence regions. As aforementioned, this behaviour is more evident when the BAO data is included as can be seen in the \Cref{fig:LDDM_decay}, with a higher best-fit $H_0^*$ value and larger fractions of relativistic daughter particle reaching $\sim 14\%$ at $95\%$ C.L., in comparison to the $\sim 8\%$ using SN+SL data alone. {This points to decaying dark matter scenario with lower values of early-time dark matter density quickly decaying with slightly larger fractions of relativistic massless daughter particle, in comparison to the SN+SL data alone.} As can be seen by contrasting the \textit{Top} and \textit{Bottom} rows of \Cref{fig:LDDM_decay_F}, the profile likelihoods when including the BAO data show larger variation from the MCMC based confidence regions\footnote{This is the expected behaviour of extremely non-Gaussian likelihoods, for example, as was described in \citet{Strege14}.}. As can be seen from both the profile likelihood and marginalised confidence regions, with the SN+SL dataset the best-fit of both the Two-body and Many-body scenarios is pushed towards large life-times and very low fraction of the relativistic daughter particle (i.e, upper-left region of the figures), reaching the limits of assumed priors in the analysis. This makes the $\Lambda$DDM model equivalent to the $\Lambda$CDM model and accordingly no improvement is found even in the $\chi^2_{\rm b.f}$ comparison. 

For the profile likelihoods obtained including the BAO data, the excluded region is reduced, in accordance with the inference also made from the marginalised confidence regions in \Cref{fig:LDDM_decay}. This is solely due to the shift in the best-fit between the SN+SL and  SN+SL+BAO data, and might appear at face-value that the constraints have become less stringent. In contrast to the SN+SL dataset, now the best-fit of the analyses prefer lower life-time and larger fractions of relativistic daughter particle ($\sim 10-15\%$). Clearly implying the effect of BAO data in combination with SL data to accommodate larger values of $H_0$, which is able to distinguish the decay dark matter from standard $\Lambda$CDM. However, the two-body decay scenario with the best-fit values of decay parameters $\{\eps, \tauG\} = \{-1.3,-2.9\}$ is statistically indistinguishable w.r.t $\Lambda$CDM having a $\Delta \chi^2_{\rm b.f} = -1.3$, and would be at a disadvantage with any information criteria, when accounting for the two additional parameters. Note that the best-fit model here does not correspond to the best-fit of $\Lambda$CDM model and hence the $\Delta \chi^2$ levels should not to be immediately contrasted with $\Lambda$CDM expectations. At the face value, the inclusion of BAO data also indicates an upper limit on the life-times of the parent decay particle, which is however only a consequence of setting the $\Delta \chi^2$ cuts and one should refer only to the marginalised confidence regions, which do not show such upper limits.

\subsection{Comment on $H_0$} 
\label{sec:Results_H0}
Note that the $\Omega_{\rm DM}^{*}$ constraints reported in \Cref{tab:LDDM_params} correspond to the early-time $\Lambda$CDM and should not to be mistaken for the final decayed dark matter density at late-times. Essentially, pointing out that in the $\Lambda$DDM framework the early time matter density should in itself reduces to give rise to higher values of late-time $H_0$, which is in fact the standard correlation also in the $\Lambda$CDM scenario. Further aided by the decay of the dark matter density the $\Lambda$DDM will yield lower values of $H_0$ than in the standard $\Lambda$CDM model. The extrapolated $H_0$ in the decaying dark matter scenarios is very much similar to the $H_0^*$ reported in \Cref{tab:LDDM_params}, with mildly larger lower error.

\begin{figure}
    \centering
    \includegraphics[scale=0.45]{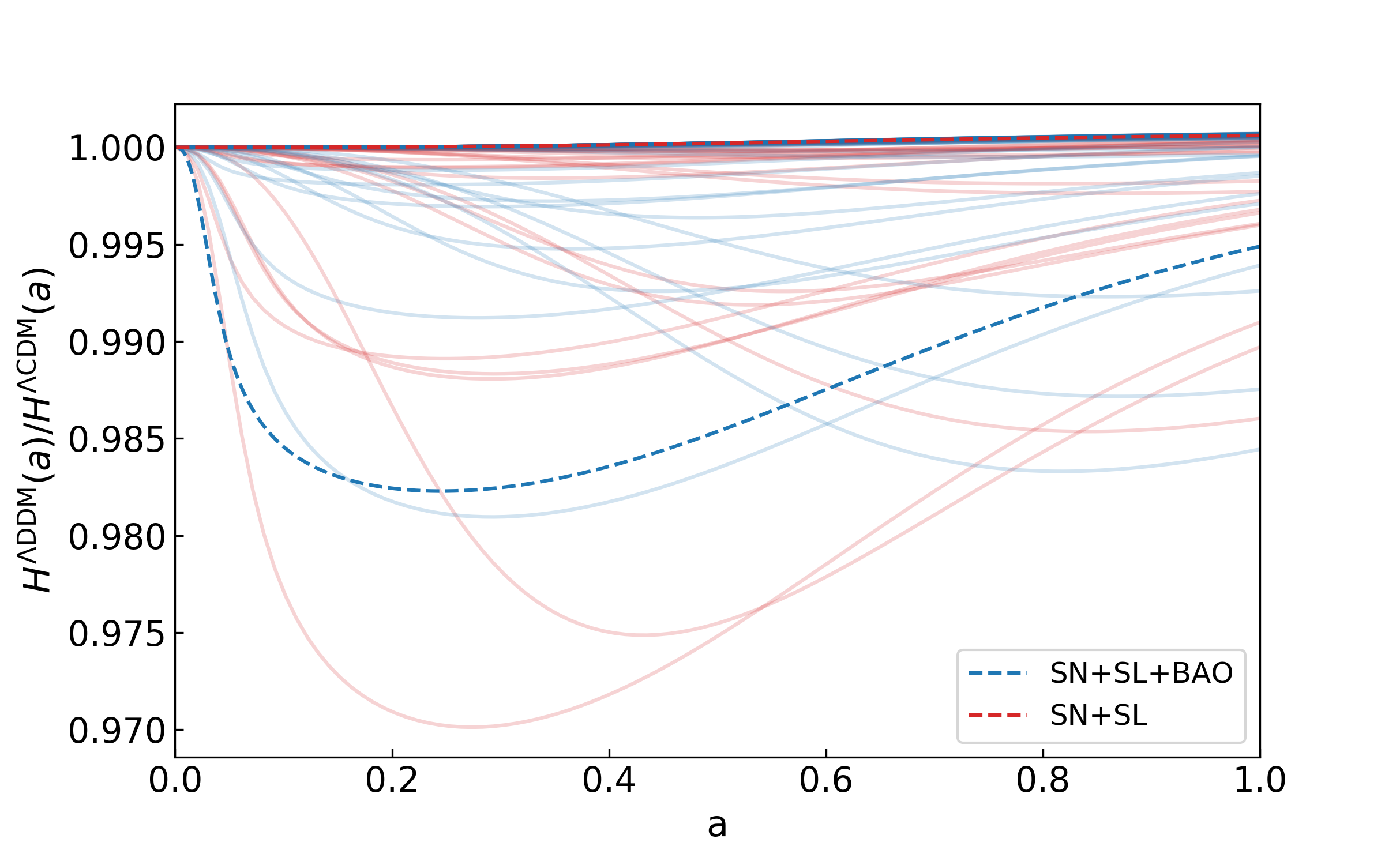}
    \caption{Here we show a 100 randomly chosen expansion histories from a larger MCMC sample of two-body $\Lambda$DDM background histories normalised to the corresponding early-time $\Lambda$CDM model, from which the late-time decaying behaviour deviates. In blue we show 50 curves for SN+SL+BAO dataset and in red we show the SN+SL data combination. The dashed curve corresponds to the best-fit model in each case reported in \Cref{tab:LDDM_params}. For the SN+SL data the best-fit coincides with $\Lambda$CDM model.}
    \label{fig:LDDMBG}
\end{figure}

We now comment on the phenomenological aspects recovered in the MCMC analysis of the current decaying dark matter model proposed to resolve the $H_0$-tension. The authors of \citetalias{Vattis19}, propose that a $\Lambda$CDM like early universe with a late-time decay of the dark matter particle, for a certain combination of decay parameters can raise the $H_0$ value in comparison to the $\Lambda$CDM expectation, potentially alleviating the $H_0$-tension. While the proposal was explicitly made with emphasis on the warm behaviour of the massive daughter particle in two-body decay scenario, we find that the same can be inferred also for the many-body decay when including the BAO data. However, we find that the decaying dark matter model, within the given parameter space mostly remains equivalent to $\Lambda$CDM and does not increase the value of the extrapolated $H_0$ in construction, through the modifications of late-time dynamics.

In \Cref{fig:LDDMBG}, we show through a few samples randomly drawn from the MCMC analysis with SN+SL (red) and SN+SL+BAO (blue) of two-body decay scenario, that the recovered $H_0$ in the $\Lambda$DDM model will remain lower than the corresponding early $\Lambda$CDM model. For the data combination of SN+SL we indeed find that the posteriors remain almost equivalent to the standard $\Lambda$CDM model\footnote{We also validate that the numerical error accumulated due to the iterative solving is only of the order of $\sim 0.05\%$ in the extrapolated $H_0$, which is clearly negligible in comparison to the statistical dispersion.}, including the best-fit model, with mild increase in dispersion towards the lower values of $H_0$. The best-fit $\Lambda$DDM model with the inclusion of BAOs indeed has a higher-value of $H_0^*$ in comparison to the $\Lambda$CDM value, however it also results in a lower value of $H_0$. While the $\Lambda$DDM model tends to grow faster in rate than the corresponding $\Lambda$CDM model fixed at early times, given that a fraction of the early cold dark matter density has decayed in the form of relativistic daughter particle at late times, the overall expansion rate remains lower than the $\Lambda$CDM expectation. We find that this mildly higher value of $H_0^*$ at $a_*$ when reconstructed to the $a = 1$ reduces to $H_0 = 69.3\, \ksM$, which is high compared to the best-fit value of $\Lambda$CDM but within its $1\sigma$ uncertainty distribution, wherein one can immediately infer that the current decaying dark matter scenarios do not perform any better to alleviate the $H_0$-tension. While our analysis is mostly consistent with the earlier analysis in Refs. \citetalias{Blackadder14, Vattis19}, in contrast, we find the inferences for lifting the $H_0$-tension to be less feasible. One might in-turn suspect that an extension of dark energy equation of state $w\neq-1$ to the decaying dark matter might tend to decrease higher values (w.r.t $\Lambda$CDM) of  $H_0$ by producing an even higher value of $H_0^*$ at recombination. However, this would also imply that the initial dark matter density has to be very low when breaking the degeneracy between $\Omega_{\rm DM}^*$ and $H_0^*$, bringing the final constraints close to the $w$CDM model with no additional effect of the decaying dark matter. We verify this by performing a simple MCMC analysis with $w$ as a free parameter, in addition to the decay parameters.

On the other hand, we find that the $\Lambda$DDM model could potentially explain the not yet significant but interesting trend of decreasing $H_0$ against increasing lens redshift ($z_d$) observed in the Strong lenses dataset \citep{Wong19, Birrer19}. This trend in $H_0$ has been observed while marginalising on the matter density, also within extended models with dark energy equation of state ($w\neq-1$) and curvature ($\Omega_{\rm k} \neq 0$). As is very well known the individual strong lensing systems are unable to constrain the matter density, which is also uncorrelated with the $H_0$, implying that the trend could not be explained in the standard scenario, if it becomes statistically significant. {Recently, in \citet{Krishnan20} a similar trend was reported with binned BAO and SN dataset, which however cannot be immediately contrasted with trend in SL dataset, as the SL system takes into account both the lens and source redshifts. We suspect that the trend in the binned data could only be a manifestation of breaking the degeneracy between $H_0$ and $\Omega_{\rm m}$ (see Table III of \citet{Krishnan20}) at different redshifts. As it is very well known, at least two data points at two distinct redshifts are required to effectively constrain the slope of expansion history and hence dark energy density. Also, one can easily verify that data (e.g., SN compilation) at redshifts lower than the deceleration-acceleration transition will yield higher dark energy density, in comparison to data at higher redshifts.}

\begin{figure}
    \centering
    \includegraphics[scale=0.45]{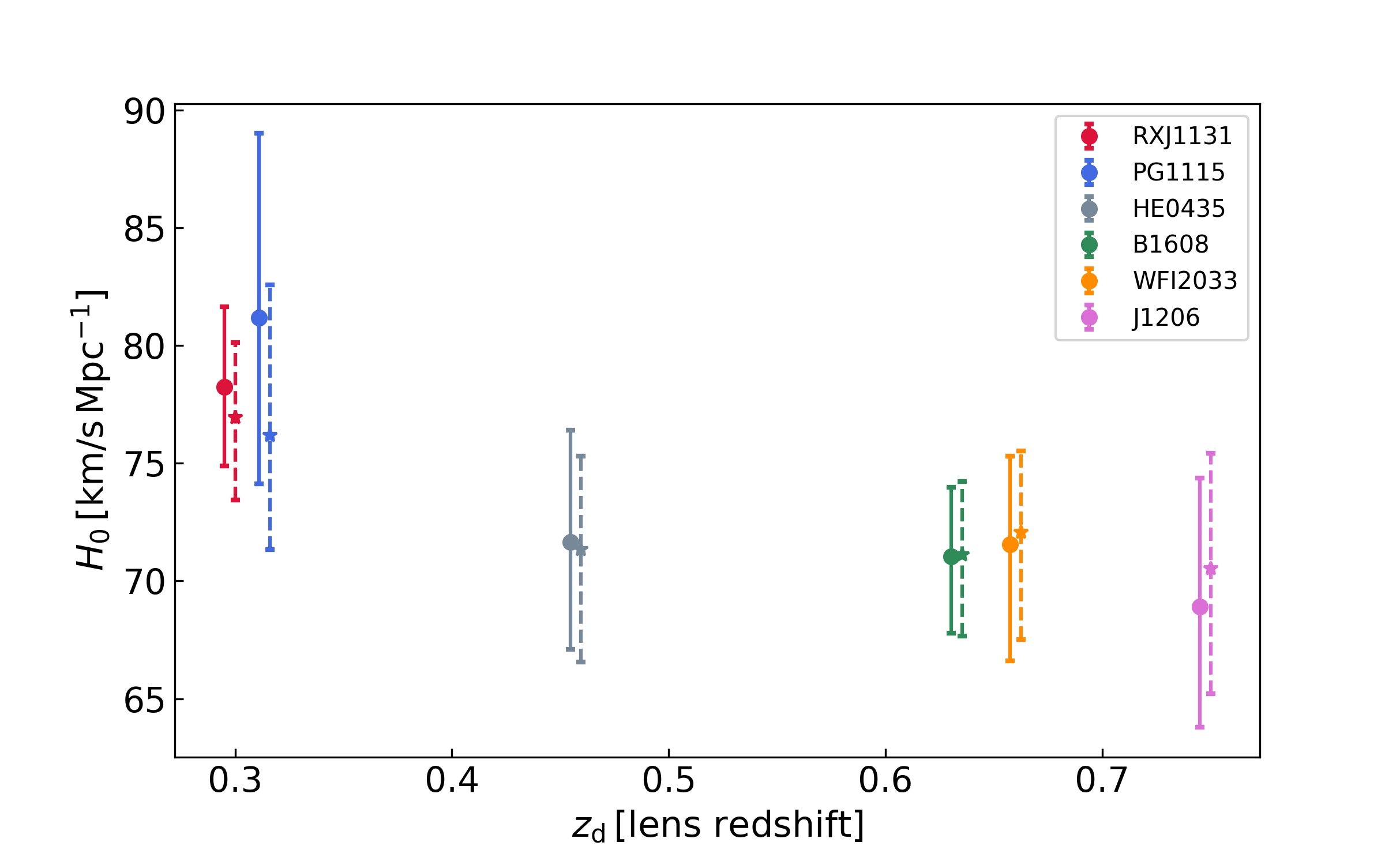}
    \caption{Comparison of constrains on $H_0$ from individual H0LiCOW lenses. The dashed error bars show our $\Lambda$DDM constrains with fixed decay parameters and the solid error bars are taken from the standard analysis in \citet{Wong19}. The dashed $H_0$ estimates in the $\Lambda$DDM scenario indicated reduced trend.}
    \label{fig:H0_Cosmo}
\end{figure}

To assess this redshift dependent trend within the $\Lambda$DDM model, we first perform the joint analysis of all six lensing systems, imposing the early universe priors, finding no constrains on the decay parameters. The best-fit however shows similar behaviour as with the inclusion of BAO data, with $\{H_0^*,\Omega_{\rm DM}^*\} = \{77.3 \, \ksM, 0.2\}$ and the decay parameters $\{ \tauG,\eps\} = \{ -0.03, -4.86 \}$. We then fix the decay parameters to their best-fit values from the joint analysis and sample on rest of the parameters in the MCMC analysis. In \Cref{fig:H0_Cosmo}, we show the comparison of the individual $H_0$ estimates from standard analysis, as in \citet{Wong_2016} (solid) and the $\Lambda$DDM scenario (dashed), with the fixed decay parameters. As suspected, we find in the $\Lambda$DDM scenario that the larger values of the $H_0$ at low-redshift are reduced and the lower values at high-redshift mildly increased. This brings all the individual estimates closer to the mean value form the joint analysis and reduces the variation in redshift. Note that best-fit lifetime that we have fixed is one particular decay scenario at early-time decay and with a larger relativistic fraction. If the current trend should become more significant with future data such as the Large Synoptic Survey Telescope (LSST) \citep{Ivezic:2008fe}, the decaying dark matter scenario can provide a suitable explanation to alleviate the trend and in turn the SL dataset will be able to place constraints on the decay characteristics. As this trend is not yet very significant we leave the analysis here as an illustration of the effect, without marginalising on the decay parameters for each of the SL data. 

\section{Conclusions}
\label{sec:Conclusions}

We present new constraints on the decaying dark matter model developed in \citetalias{Blackadder14}, allowing for a possibly warm massive daughter particle performing MCMC analysis. We update the bounds on the life-time of the decay particle to be $\tau > 9 \, \rm{Gyr}$ and $\tau > 11 \, \rm{Gyr}$, for the two-body and many-body decay scenarios at $95\%$ C.L., for maximum allowed relativistic massless fraction in each of the cases. Our limits are mildly less stringent than the limits earlier reported in \citetalias{Blackadder14} and we are unable to place limits on life-time for a $1\%$ relativistic fraction as were presented in these previous works. Our inference of $\tau > 59 \, \rm{Gyr}$ at $95\%$ C.L., for a fixed $H_0$ is an improvement and is in very good agreement with the earlier bounds placed in \citet{Aubourg14}, appropriately comparing within the larger parameter space available in our analysis. By comparing our primary marginalised confidence regions with profile likelihoods, we highlight the importance of not fixing the background parameters when obtaining the limits on the decay life-time. 

Alongside updating the bounds, we assessed the feasibility of the late-time decaying dark matter model proposed as a possible late-time resolution for the $H_0$-tension. While we validate the claim of \citetalias{Vattis19}, we also find that the decaying dark matter resolution in effect might not be feasible to resolve the $H_0$-tension, owing to the very mild increase in comparison to the $\Lambda$CDM model. We find that the current decaying dark matter scenario is able to alleviate the mild trend \citep{Wong19} observed for the decreasing $H_0$ estimates with increasing lens redshift in the strong lenses dataset. While this trend is not yet statistical significant, the $\Lambda$DDM model would be an appropriate alternative if the future strong lensing datasets (LSST \citet{Ivezic:2008fe}), were to strengthen the trend. 

The decaying dark matter model provides interesting scenarios and such a late-time variation of the physics from the standard $\Lambda$CDM model must be further investigated, also in light of upcoming low-redshift experiments like Euclid \citep{Amendola16}, DESI \citep{Levi13} etc., which can potentially constrain extensions of the standard scenario to unprecedented accuracy.

\section*{Acknowledgements}

B.S.H acknowledge financial support by ASI Grant No. 2016-24-H.0. M.V. is supported by INFN INDARK PD51 grant and agreement ASI-INAF n.2017-14-H.0. We acknowledge the use of CINECA high performance computing resources under the projects `INF19\_indark\_0', `INF20\_indark' and Marina Migliaccio for help with the same. We thank the authors of \citet{Vattis19}, for an early correspondence providing clarifications on their model. {We thank Julien Lesgourgues, Vivian Poulin and Riccardo Murgia for useful comments on the draft. }




\bibliographystyle{mnras}
\bibliography{bibliografia} 







\bsp	
\label{lastpage}
\end{document}